\documentclass[conference]{IEEEtran}
\IEEEoverridecommandlockouts
\usepackage{cite}
\usepackage{amsmath,amssymb,amsfonts}
\usepackage{algorithmic}
\usepackage{graphicx}
\usepackage{textcomp}
\usepackage{xcolor}
\usepackage{siunitx}
\def\BibTeX{{\rm B\kern-.05em{\sc i\kern-.025em b}\kern-.08em
    T\kern-.1667em\lower.7ex\hbox{E}\kern-.125emX}}

\newcommand\eu{\ensuremath{\mathrm{e}}}
\newcommand\fmax{\ensuremath{f_{\mathit{max}}}}
\begin{document}

\title{Broadband MEMS Microphone Arrays with Reduced Aperture Through 3D-Printed Waveguides}

\author{
\IEEEauthorblockN{Dennis Laurijssen\IEEEauthorrefmark{1}\IEEEauthorrefmark{2}, 
Walter Daems\IEEEauthorrefmark{1}\IEEEauthorrefmark{2}, 
Jan Steckel\IEEEauthorrefmark{1}\IEEEauthorrefmark{2}}
 \IEEEauthorblockA{\IEEEauthorrefmark{1}Cosys-Lab, Faculty of Applied Engineering, University of Antwerp, Antwerp, Belgium}
 \IEEEauthorblockA{\IEEEauthorrefmark{2}Flanders Make Strategic Research Centre, Lommel, Belgium\\
 \IEEEauthorrefmark{1}dennis.laurijssen@uantwerpen.be}
}

\maketitle

\begin{abstract}
In this paper we present a passive and cost-effective method for increasing the frequency range of ultrasound MEMS microphone arrays when using beamforming techniques. By applying a 3D-printed construction that reduces the acoustic aperture of the MEMS microphones we can create a regularly spaced microphone array layout with much smaller inter-element spacing than could be accomplished on a printed circuit board due to the physical size of the MEMS elements. This method allows the use of ultrasound sensors incorporating microphone arrays in combination with beamforming techniques without aliases due to grating lobes in applications such as sound source localization or the emulation of bat HRTFs.
\end{abstract}


\section{Introduction}
In recent years, in-air ultrasound technology has garnered significant advances due to the development of ultrasound MEMS microphones making it more feasible to build (large) arrays at low cost. Active in-air ultrasound, which involves the emission and reception of sound waves at frequencies beyond 20\si{\kilo\hertz}, better known as sonar, can be applied for non-contact interaction, distance measurements, and the ability to penetrate certain materials due to its wavelength. These characteristics make it an attractive option for enhancing robotic sensing and control systems\cite{8794165,6331017,jansen2024semantic,jansen2022real,hernandez2017fpga}.

A critical component in in-air ultrasound systems is the microphone array, which is used for receiving the impinging sound waves, and consists of multiple microphones arranged in a specific geometric pattern. The performance of these arrays in combination with beamforming techniques is highly dependent on the inter-element spacing between adjacent microphones. The inter-element spacing determines the maximum frequency that can be used for beamforming without introducing grating lobes in the directivity pattern. These grating lobes are unwanted spatial aliases that occur besides the main lobe when the spacing between array elements exceeds half the wavelength of the ultrasound signal. These lobes result in false peaks in the array's directional response, leading to ambiguity in sound source localization. In the context of robotics, grating lobes can severely degrade the performance of ultrasonic sensors, causing erroneous detection of echos and unreliable environmental mapping. This can be particularly detrimental when these robotic systems rely heavily on accurate and precise sensory information for navigation, object detection, and interaction with the environment, thus compromising the robot's ability to perform tasks efficiently and safely.

To address this issue, we propose a solution involving the reduction of the inter-element spacing of the microphones using a 3D-printed construction reduces the acoustic aperture of the microphones. By carefully designing and implementing these so-called baffles to match the microphone array on the PCB, we can achieve closer spacing between the inlets of the baffles than we could with the microphones without the baffles, which mitigates the formation of grating lobes. This innovation ensures a higher frequency bandwidth while beamforming, and significantly enhances the reliability of ultrasonic sensing in various applications. The 3D-printed baffles allow for a cost-effective approach in customizing the microphone layout, that would otherwise be impossible due to the physical size of the MEMS microphone elements. While applying 3D-printed baffles to alter directivity patterns in ultrasound applications has been previously applied on piezoelectric transducer~\cite{rutsch2021waveguide,9593561,9126862,7792174,10324935}, we now transfer this approach to a receiver array of MEMS microphones.

This paper explores the design and configuration of microphone arrays, and the implications of inter-element spacing on beamforming performance through both simulation and experimentation. It shows the adverse effects of grating lobes and details our proposed solution of using a 3D-printed construction which incorporates baffles to reduce inter-element spacing. Through this analysis, the paper aims to contribute to the advancement of ultrasonic sensing technologies in robotics, enhancing their robustness and applicability in various operational environments.

\section{Measurement System}
When designing ultrasound sensors that support sound source localization, microphone arrays are commonly used in combination with beamforming. These array configurations can vary from circular, random\cite{kerstens2019, kerstens2017low}, spiral\cite{allevato2022air,allevato2023two,8720363}, and 1D and 2D\cite{verellen2020urtis,10491247, verellen2020} regularly spaced layouts. When utilizing the latter array configuration, the Nyquist-Shannon theorem can be applied to the spatial domain\cite{dmochowski2008spatial} as described in equation \ref{eq:nyquistspat}. The inter-element spacing $d$ and the speed of sound in air $\mathit{v}$, which we approximate to be 343\unit[per-mode = symbol]{\meter\per\second}, is used to determine at what frequency \fmax{} grating lobes are introduced besides the main lobe when beam steering\cite{konetzke2015phased}. These grating lobes manifest themselves as spatial aliases causing false detections of echos of transmitted ultrasound or ultrasound sources.

\begin{align} 
d = \frac{\lambda}{2} = \frac{v}{2f}  
\quad\Leftrightarrow\quad
f = \frac{v}{2 d}
\label{eq:nyquistspat}
\end{align}

By applying (\ref{eq:nyquistspat}) to our measurement system, a custom developed ultrasound sensor called eRTIS~\cite{kerstens2019}, equipped with a 2D regularly spaced MEMS microphone array with a spacing of 3.8\si{\milli\meter}, we can determine that the frequency \fmax{} at which grating lobes are introduced into the directivity pattern is 45.13\si{\kilo\hertz}. The transducer on the aforementioned eRTIS sensor however is capable of emitting ultrasonic calls up to 100\si{\kilo\hertz}, well beyond the \fmax{} of the microphone array.

\subsection{3D Printed Baffles to Reduce the Acoustic Aperture}
\begin{figure}
    \centering
    \includegraphics[width=0.85\linewidth]{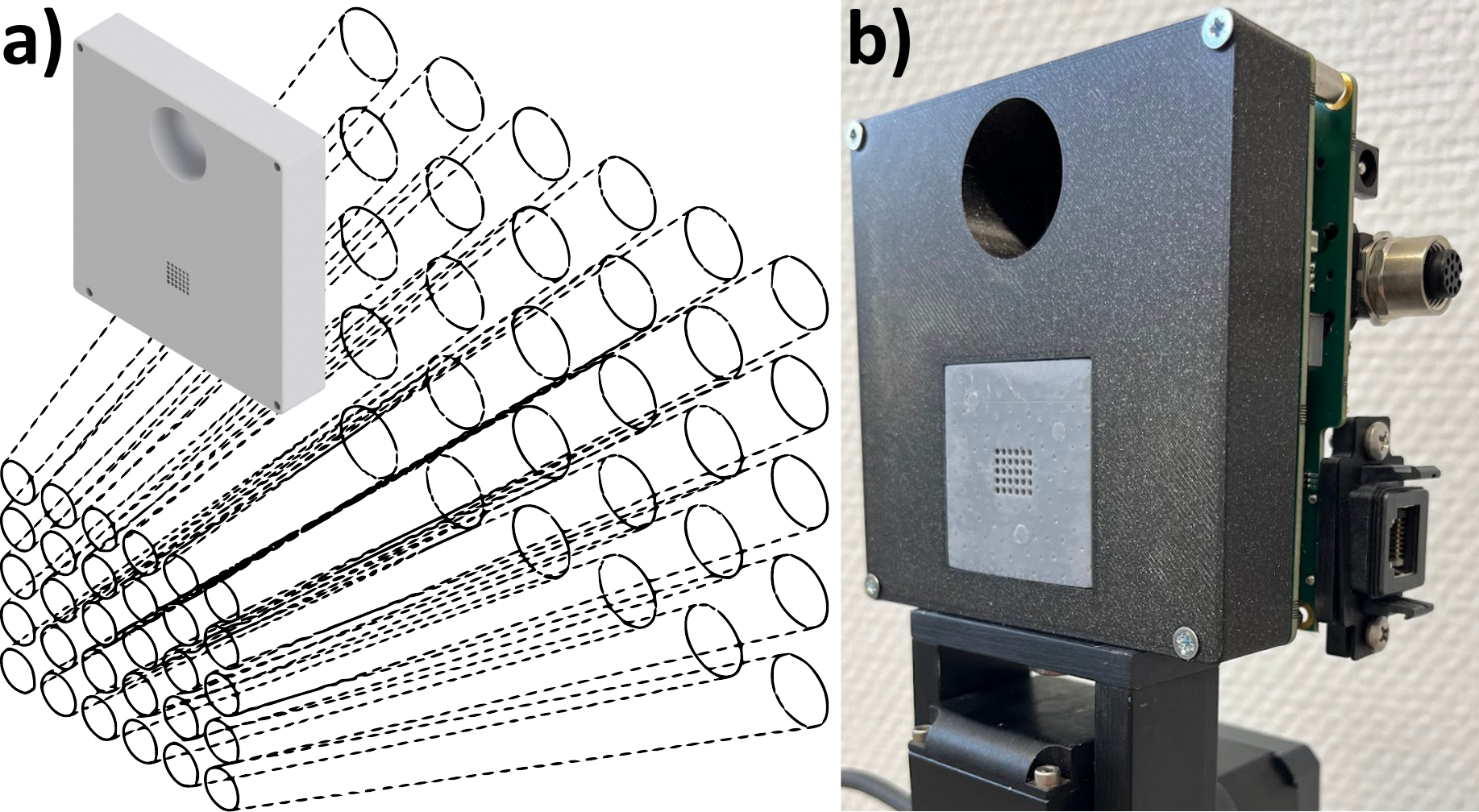}
    \caption{a) Wireframe representation of the 30 waveguides inset, incorporated into the 3D-printed baffle structure of which the render is also shown in the top left corner, that reduce the inter-element spacing of the MEMS microphones from 3.8\si{\milli\meter} to 1.8\si{\milli\meter}. This reduced inter-element microphone array layout effectively increases the \fmax{} at which grating lobes would be introduced in the directivity pattern from 45.13\si{\kilo\hertz} to 95.25\si{\kilo\hertz}. b) Shows the fabricated 3D-baffle structure with the SLA-printed  inset and the outer FDM-printed assembled with the eRTIS ultrasound sensor.}
    \label{fig:baffle}
\end{figure}

Given that it is unfeasible to further reduce the spacing between the MEMS microphones in the array layout due to the size of these components, another approach was developed to further decrease the inter-element spacing. This approach relies on using a 3D printed add-on structure that was designed to reduce the inter-elements spacing of the microphone's acoustic ports from 3.8\si{\milli\meter} to 1.8\si{\milli\meter}, effectively increasing the \fmax{} of the array to 95.25\si{\kilo\hertz} before grating lobes would be introduced to the directivity pattern when using beamforming. This add-on structure, also referred to as 3D-printed baffle, is composed of 30 waveguides that allow the acoustic energy from impinging ultrasonic sound waves to reach the acoustic port of the MEMS microphones. The design of these waveguides is very straightforward, and was intentionally not optimized for facilitating the fabrication process, were we lofted the holes of the acoustic ports in the PCB to the accompanying holes to the front of the baffle structure as can be seen in figure~\ref{fig:baffle}.

To transfer this CAD design to the real world, a masked stereolithography (MSLA) 3D-printer was used (Elegoo Mars 2 Pro). Whereas Fused Deposition Modeling (FDM) 3D-printers create objects of 3D CAD models by extruding thermoplastic filaments through a heated nozzle, which melts the material and deposits it layer by layer onto a build platform, an MSLA printer uses ultraviolet light to cure photosensitive polymer resins one layer at a time on a build plate. While generally this is a slower process, it allows for high-accuracy, isotropic parts that require much tighter tolerances. While the chosen SLA technology is capable of realizing our design in great detail, we increased the diameter of the waveguides to make sure the resin can exit the cavities despite surface tension. Due to the size of the entire part covering the front-end of the ultrasound sensor (measuring 10\si{\centi\meter} by 10\si{\centi\meter}), it would not fit inside the build volume of the SLA-printer. Therefore, the 3D-printed structure was split up in an inset that would fit the inside the SLA printer's build volume, and an outer part that was created using an FDM 3D-printer.

\section{Experimental Setup and Results}
Once our 3D-printed baffle was fabricated, we mounted the baffle to the front of the eRTIS device making sure the gap between the front of the PCB and the back of the 3D-printed baffle would be as small as possible by using some tape in between. The 3D-printed baffle also contains a 1/4 inch threaded insert, that is aligned to the vertical center of the microphone array, in order to mount it to a FLIR PTU-46 pan tilt system. A Senscomp 7000 transducer is placed perpendicular in front of the sensor when in a neutral position with $\theta = 0$ at a range of approximately 2\si{\meter}. This setup was placed in a large environment to minimize reflections. The pan angle $\theta$ was varied between $-90\si{\degree}$ to $+90\si{\degree}$ with $1\si{\degree}$ steps performing 10 measurements per step. For these measurements, a logarithmic FM sweep of 2.5\si{\milli\second} was emitted from 100\si{\kilo\hertz} down to 20\si{\kilo\hertz} by the Senscomp transducer which was recorded by the eRTIS device and transmitted using a USB connection to a PC for storage and further processing. 

\subsection{Baffle Transfer-Function Calibration}
Given that the waveguides in the 3D-printed baffle construct are not equal in length, phase shifts between the microphones are introduced when measuring the impinging sound waves. In order to compensate for the phase shifts that are introduced by applying this approach, a calibration step is necessary to adjust the input signal $s_{M_{i}}(t)$ to the phase calibrated input signal $s_{M_{c_{i}}}(t)$ with $i$ the index of microphone $M$ in the array. In order to do this we apply the following reasoning:
\begin{align} 
s_{M_{i}}(t) &= h_i(t)*s(t-\Delta t_i)\\
             &\downarrow \mathcal{F}\nonumber\\
S_{M_{i}}(\omega) &= H_i(\omega)\cdot S(\omega) \cdot \eu^{-j \omega \Delta t_i} \label{eq:calibp1}
\end{align}
with $h_i(t)$ the filter kernel describing the alteration of the emitted sound on its baffled path from the emitter to microphone $i$ (without taking the delay into account due to the geometry of the array).\\
Then we estimate $H_i(\omega)$ using the measurement for $\theta = 0$, i.e. $\Delta t_i = 0$, such that
\begin{align}
  S_{M_{i}}(\omega) = H_i(\omega) \cdot S(\omega)  \Leftrightarrow  H_i(\omega) = \frac{S_{M_{i}}(\omega)}{S(\omega)}
\end{align}
Subsequently, we calibrate (\ref{eq:calibp1}) with $H_{f_{i}}(\omega)=\left(\frac{H_i(j\omega)}{|H_i(j \omega)|}\right )^*\nonumber$:
\begin{align}
S_{M_{c_{i}}}(\omega)&=H_{f_{i}}(\omega) \cdot H_i(\omega) \cdot S(\omega)\cdot\eu^{-j \omega \Delta t_i} \xrightarrow{\mathcal{F}^{-1}} s_{M_{c_{i}}}(t)
\end{align}
in which we approximate $S(\omega) \approx \frac{1}{N}\sum_{i=1}^{N}S_{M_{i}}(\omega)$.

\subsection{Experimental Results}
\begin{figure}
    \centering
    \includegraphics[width=0.85\linewidth]{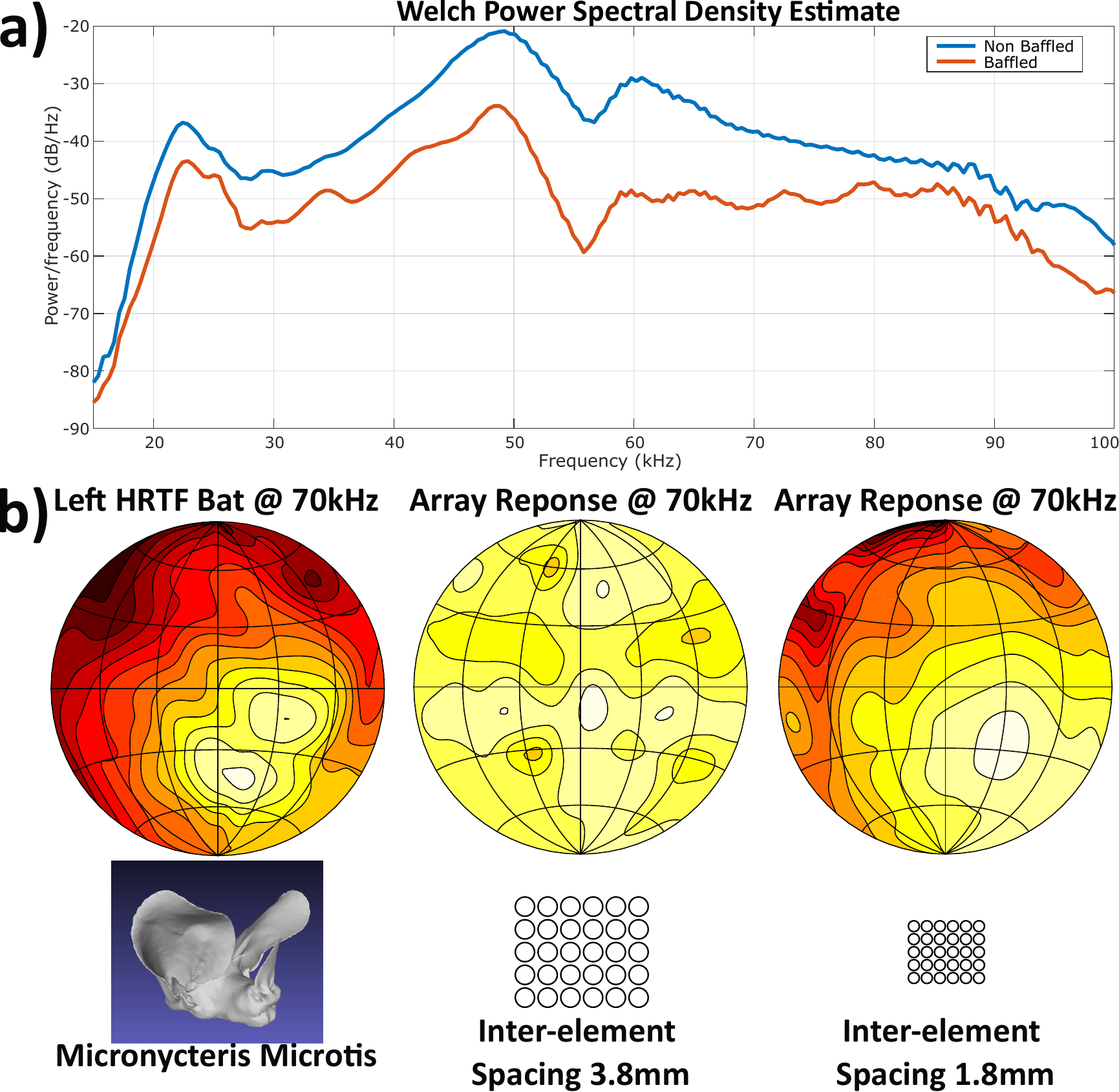}
    \caption{a) The Welch Power Spectral Density is shown comparing both the recorded non-baffled recordings with the baffled recordings, in which the blue curve shows the non-baffled PSD and the red curve shows the baffled PSD. This shows that by using the 3D-printed baffle add-on in front of the microphone the signal gets attenuated by a maximum of approximately 15\si{\decibel}. b) Shows the HRTF of the left ear of a Micronycteris Microtis bat and the synthetic HRTF approximations of a microphone array layout with an inter-element spacing of 3.8\si{\milli\meter} and 1.8\si{\milli\meter}}
    \label{fig:psd}
\end{figure}

In order to compare the recorded data with a baseline, a similar setup was used where the 3D-printed baffle was replaced with a PCB holder respecting the same vertical center line to rotate around. The same pan angles, number of measurements per pan angle and FM sweep were used to record a non-baffled dataset. As an initial comparison, both datasets were averaged out per pan angle for which the Welch Power Spectral Density was calculated. The results of these PSDs are shown in figure~\ref{fig:psd} comparing both the recorded non-baffled recordings with the baffled recordings, in which the blue curve shows the non-baffled PSD and the red curve shows the baffled PSD. This shows that by using the 3D-printed baffle add-on in front of the microphone the signal gets attenuated by a maximum of approximately 15\si{\decibel}. 

\begin{figure}
    \centering
    \includegraphics[width=0.85\linewidth]{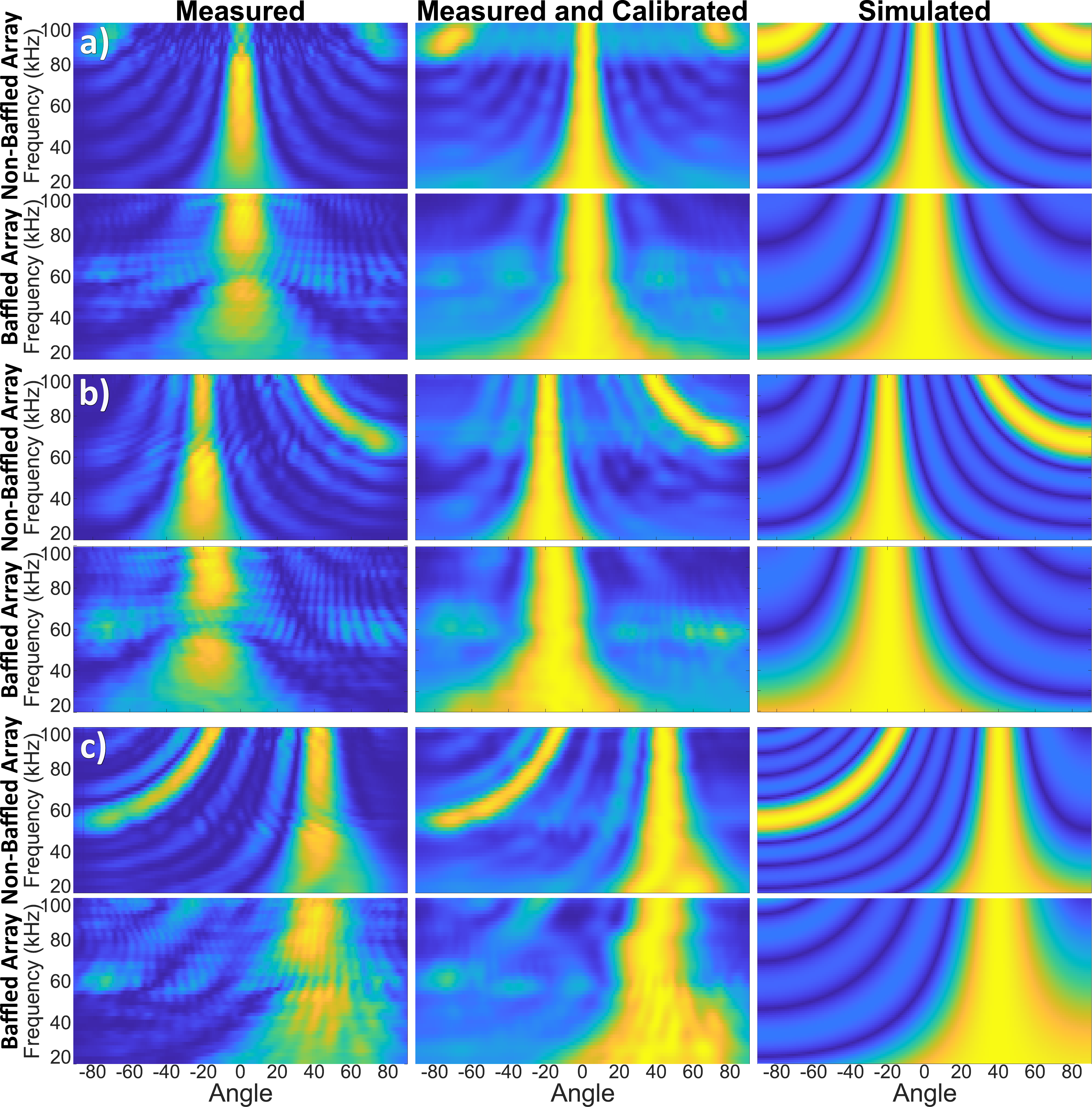}
    \caption{The effect of using baffles on the directivity pattern is shown. The columns represent the raw measured data, the calibrated measured data and simulated data with the uneven rows being the non-baffled array setup and the even rows the baffled array setup. In a) the Senscomp 7000 transducer has a pan angle of $0\si{\degree}$ in relation to the microphone array, in b) the pan angle is $-20\si{\degree}$ and in c) the pan angle is $40\si{\degree}$.}
    \label{fig:directivitypattern}
\end{figure}

Subsequently, the effect of the baffles on the directivity pattern when using beamforming is shown in figure~\ref{fig:directivitypattern}. In this figure the plots in the columns represent the raw measured data, the calibrated measured data and simulated data with the uneven rows corresponding to the non-baffled array setup and the even rows to the baffled array setup. In a) the Senscomp 7000 transducer has a pan angle of $0\si{\degree}$ in relation to the microphone array, in b) the pan angle is $-20\si{\degree}$ and in c) the pan angle is $40\si{\degree}$. The grating lobes introduced in the plots of the non-baffled setup are clearly visible in the frequency ranges above 45\si{\kilo\hertz} in both the simulated and the measured data whereas these bands are eliminated in the baffled array setup.

\section{Discussion and Conclusion}
In this paper, we propose and demonstrate how a cheap 3D-printed baffle add-on structure can be used in combination with microphone arrays to increase the frequency range of in-air ultrasound measurement systems before grating lobes are introduced in the directivity pattern when beamforming. While we observed a considerable amount of attenuation in the signal by using this approach, we believe this is could be mitigated by using optimized waveguide shapes instead of cylinders. As shown in the previous section, a calibration step is necessary to compensate for the phase shifts that are introduced by the length of the waveguides after which the results of the measurements match the simulations to a great extent.

Given that the physical size of the MEMS microphones is not likely to shrink even further, we believe that this baffled approach is viable option for various applications, e.g., in robotics where less post-processing~\cite{9251495,9931740} steps are necessary to negate the spatial aliases in the acoustic images as a result of the grating lobes in the directivity pattern of the beamformer. We also envision applications where we could better mimic the HRTF (Head Related Transfer Function) of bats~\cite{6242793, vanderelst2010noseleaves} as shown in figure~\ref{fig:psd}b), of which some species are known to use ultrasonic calls up to 180\si{\kilo\hertz}, to investigate their foraging behaviour.

\clearpage
\newpage

\bibliographystyle{IEEEtran}
\bibliography{citationsLibrary}

\end{document}